\documentclass{article}

\usepackage[usenames,dvipsnames]{xcolor}

\usepackage{cite}
\usepackage{caption}
\usepackage{graphicx}
\usepackage{epsfig}
\usepackage{amsmath,amssymb,bm,stmaryrd}
\usepackage{delarray}
\usepackage{stfloats}
\usepackage{float}
\usepackage{multirow,hhline}

\usepackage{algorithm}
\usepackage{algorithmic}
\usepackage{arydshln}
\usepackage{cuted}
\usepackage{balance}
\usepackage{verbatim}
\usepackage{array}
\usepackage{authblk}
\newcolumntype{L}{>{\centering\arraybackslash}m{2.2cm}}

\usepackage{etoolbox}

\makeatletter
\patchcmd{\@maketitle}{\raggedright}{\centering}{}{}
\patchcmd{\@maketitle}{\raggedright}{\centering}{}{}
\makeatother

\begin{document}

\title{\begin{LARGE}\textit{In Vivo} WBAN Communication: Design and Implementation\end{LARGE}}

\author{Hadeel Elayan$^1$, Raed Shubair$^2$, and Nawaf Almoosa$^1$}
\affil{$^1$Electrical and Computer Engineering Department, Khalifa University, UAE}
\affil{$^2$Research Laboratory of Electronics, MIT}
\affil{\small Email: hadeel.mohammad@kustar.ac.ae, raed.shubair@kustar.ac.ae, nawaf.almoosa@kustar.ac.ae}
\date{}
\maketitle


The emerging \textit{in vivo} communication and networking system is a prospective component in advancing healthcare delivery and empowering the development of new applications and services. \textit{In vivo} communications is based on networked cyber-physical systems of embedded devices to allow rapid, correct and cost-effective responses under various conditions. This chapter presents the existing research which investigates the state of art of the \textit{in vivo} communication. It focuses on characterizing and modeling the \textit{in vivo} wireless channel and contrasting it with the other familiar channels. MIMO \textit{in vivo} is also of cencern in this chapter  since it significantly enhances the performance gain and data rates. Furthermore, this chapter addresses \textit{in vivo} nano-communication which is presented for medical applications to provide fast and accurate disease diagnosis and treatment. Such communication paradigm is capable of operating inside the human body in real time and will be of great benefit for medical monitoring and medical implant communications. Consequently, propagation at the Terahertz (THz) frequency must be well understood as it is considered the most promising band for electromagnetic  nano-communication models. 


\section{Introduction }
Wireless Body Area Networks (WBANs) are a new generation of Wireless Sensor Networks (WSNs) dedicated for healthcare monitoring applications. The aim of these applications is to ensure continuous monitoring of the patients' vital parameters, while giving them the freedom of moving thereby  resulting in an enhanced quality of healthcare \cite{5931518}. In fact, a WBAN is a network of wearable computing devices operating on, in, or around the body. It consists of a group of tiny nodes that are equipped with biomedical sensors, motion detectors, and wireless communication devices which incoporates techniques similar to those implemetated in wireless systems \cite{1632650,shubair2015vivo,shubair_robust_2005,shubair_performance_2005,khan_compact_2017,belhoul_modelling_2003,shubair_robust_2004,shubair_closed-form_1993,omar_uwb_2016,al-ardi_direction_2006,nwalozie_simple_2013,al-nuaimi_direction_2005,bakhar_eigen_2009}. Actually, advanced healthcare delivery  relies on both body surface and internal sensors since they reduce the invasiveness of a number of medical procedures\cite{6208476}. Electrocardiogram (ECG), electroencephalography (EEG), body temperature, pulse oximetry (SpO$_{2}$), and blood pressure are evolving as long-term monitoring sensors for emergency and risk patients \cite{wegmuller2007intra}.
 
One attractive feature of the emerging Internet of Things is to consider \textit{in vivo} networking for WBANs as an important application platform that facilitates continuous wirelessly-enabled healthcare \cite{7032380}. Internal health monitoring \cite{6063492}, internal drug administration \cite{5165947}, and minimally invasive surgery \cite{6090392} are examples of the pool of applications that require communication from \textit{in vivo }sensors to body surface nodes. However, the study of \textit{in vivo} wireless transmission, from inside  the body to external transceivers is still at its early stages. Fig. 1 shows a modified network organization for interconnecting the
biomedical sensors. The data is basically not directly transferred  from the biomedical sensors to the hospital infrastructure. Indeed, sensors send their data via a suitable low-power and low-rate \textit{in vivo }communication link to the central link sensor (located on the body like all other sensors). Any of the sensors may act as a relay  between the desired and the central link sensor if a direct connection is limited. An external wireless link enables the data exchange between the central link sensor and the external hospital infrastructure \cite{wegmuller2007intra}. 

\begin{figure}[h]
\centering
\includegraphics[width=0.6\textwidth,clip]{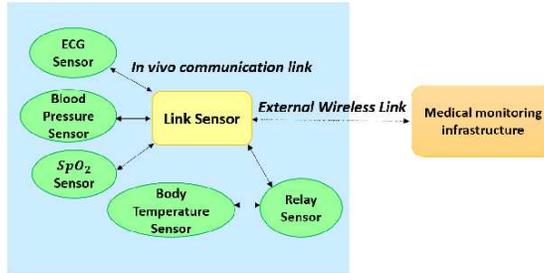}
\caption{\footnotesize Simplified overview of the \textit{in vivo} communication network.}
\label{system}
\end{figure}

Wireless \textit{in vivo} communication creates a wirelessly-networked cyber-physical system of embedded devices. Such systems utilize real-time data to enable rapid, correct as well as cost-conscious responses for surgical, diagnostic, and emergency circumstances \cite{6208476}. The crucial element  that should be carefully regarded when referring to \textit{in vivo} communications is modeling the \textit{in vivo }wireless channel. The ability to understand the characteristics of the \textit{in vivo} channel is fundamental to achieve optimum processing and design effective protocols that enable the arrangement of WBANs inside the human body \cite{6208476}.

This chapter surveys the existing research which investigates
the state-of-art of the \textit{in vivo} communication. It also focuses
on characterizing and modeling the \textit{in vivo} wireless channel
and contrasting this channel with the other familiar ones. MIMO \textit{in vivo} is also of interest since it significantly enhances the performance gain and data rates. Finally, this chapter introduces \textit{in viv}o nano-communication as a novel communication paradigm. The rest of the chapter is organized as follows. In Section II, we present the state-of-art of \textit{in vivo} communication. Conducted research on \textit{in vivo} channel characterization is provided in Section III. The MIMO \textit{in vivo} system is described in Section
IV. \textit{In vivo} nano-communication is addressed in Section V. Finally, we draw our conclusions and summarize the paper in Section VI.

\section{State-of-art of \textit{In Vivo} Communication\\}

\textit{In vivo} communication is a genuine signal transmission field which utilizes the human body as a transmission medium for electrical signals \cite{Zimmerman1995,elayan2017terahertz,elayan2017wireless}. The body becomes a vital component of
the transmission system.  Electrical current
induction into the human tissue is enabled through sophisticated transceivers while  smart data transmission is  provided  by
advanced encoding and compression.
Fig. 2 shows the main components of an \textit{in vivo} communication link. 

\begin{figure}[h]
\centering
\includegraphics[width=0.6\textwidth,clip]{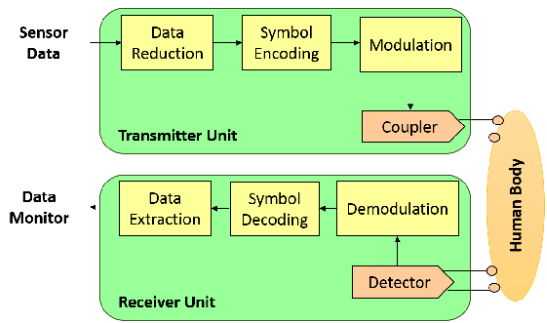}
\caption{\footnotesize\textit{In vivo} communication for data transmission between sensors enabled by transmitter and receiver units.}
\label{system}
\end{figure}

A transmitter unit permits sensor data to be compressed and encoded. It then conveys the data by a current-controlled coupler unit. The human body acts as the transmission channel. Electrical signals are coupled into the human tissue and distributed over multiple body regions. On the other hand, the receiver unit is composed of an analog detector unit that amplifies the induced signal and digital entities for data demodulation, decoding, and extraction \cite{wegmuller2007intra}.

Developing body transmission systems have shown the viability of transmitting electrical signals through the human body. Nonetheless, detailed characteristics of the human body are lacking so far. Not a lot is known about the impact of human tissue on electrical signal transmission. Actually, for advanced transceiver designs, the effects and limits of the tissue have to be cautiously taken into consideration \cite{wegmuller2007intra}\cite{668752}.
The main requirements of an \textit{in vivo} system include low power, low  latency,  less complexity, robustness to jamming, reliability, and size compactness \cite{455}.
\textit{ In vivo} communication is involved in a wide array of
practical medical usages. For instance, \textit{in vivo} sensors are
utilized in health monitoring applications in order to keep track
of glucose and blood pressure levels. \textit{In vivo} actuators are
also important for implanted insulin pumps as well as bladder
controllers. Moreover, \textit{in vivo} technology is involved in both
medical nanorobotic device communication and in therapeutic
nanoparticles employed in malignant tumor elimination processes.
Such distinctive communication can add an
effective contribution in the development of Prosthetics including artificial retina, cochlear implants and brain pacemakers for patients with Parkinsons disease.

\subsection{Human Body Model}

Research into \textit{\textit{\textit{in vivo}}} communications primarily used the ANSYS HFSS \cite{press_ansoft}  Human Body Model software to conduct the simulations. This software is a high-performance full-wave electromagnetic (EM) field simulator which enables the complete electromagnetic fields prediction and visualization. Hence, important parameters such as S-Parameters, resonant frequency, and radiation characteristics of antennas can be computed and plotted. The human body is modeled as an adult male body with more than 300 parts of muscles, bones and organs, having a geometric accuracy of 1 mm and realistic frequency dependent material parameters. The original body model only has the parameters from 10 Hz to 10 GHz.  However, the maximum operating frequency is increased to 100 GHz by manually adding the values of the parameters to the datasets \cite{gabriel1996compilation}. 

\subsection{System Level Setup}

To evaluate the Bit Error Rate (BER) performance of the \textit{in vivo} communication, an OFDM-based (IEEE 802.11n) wireless transceiver model operating at 2.4 GHz is setup. This model implies varying different
Modulation and Coding Schemes (MCS) index values as well as  bit rates in Agilent SystemVue for various \textit{in vivo }channel setups in HFSS. The system block diagram is shown in Fig. 3 \cite{6857757}.

\begin{figure}[h]
\centering
\includegraphics[width=0.6\textwidth,clip]{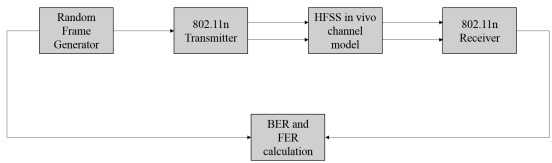}
\caption{ \footnotesize Block diagram of system level simulation with HFSS \textit{in vivo} channel model }
\label{system}
\end{figure}

\section{\textit{In Vivo} Channel Modeling and Characterization}

 The \textit{in vivo} channel is a novel paradigm in the field of wireless propagation; thus, it is very different when compared to other frequently analyzed wireless environments such as cellular,  Wireless Local Area Network (WLAN), and deep space \cite{1632650}. Fig. 4 illustrates the classic multi path channel and the \textit{in vivo }multipath channel. 

\begin{figure}[h]
\centering
\includegraphics[width=0.6\textwidth,clip]{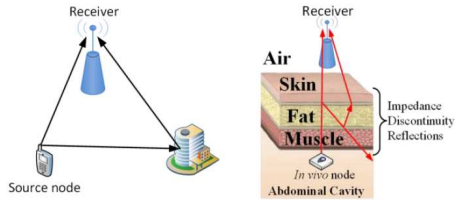}
\caption{\footnotesize Classic multi-path channel vs. \textit{in vivo} multi-path channel \cite{6208476}.}
\label{system}
\end{figure}

Basically, in an \textit{in vivo }channel, the electromagnetic wave passes through various dissimilar media that have different electrical properties, as depicted in Fig. 4. This leads to the reduction in the wave propagation speed in some organs and the  stimulation of  significant time dispersion that differs with each organ and body tissue \cite{6857757}. This is coupled with attenuation due absorption by the different layers result in the degradation of the quality of the transmitted signal in the \textit{in vivo} channel. 

The authors in \cite{wegmuller2007intra} compare the characteristics of wireless technologies including the WLAN, Bluetooth, Zig-bee, and active Radio Frequency Identification (RFID) as shown in Table 1. Their aim is 
 to seek a novel transmission technique for \textit{in vivo} communication which focuses on transmission power below 1mW, data rates of 64 kbit/s, and the possibility for miniaturization to integrate the transceiver modules into band-aids and implantable pills. 

%

\begin{table}[h]
\centering
\caption{Characteristics of Wireless Technologies \cite{wegmuller2007intra}}
\label{simulation}
\scalebox{0.9}{
\begin{tabular}{|L|L|L|L|L|}
\hline
\textbf{Technology}   & \textbf{Frequency} & \textbf{Data Rate}      & \textbf{Transmission Power} & \textbf{Size} \\ \hline
WLAN                  & 2.4/5.1 GHz        & 54 Mbit/s               & 100 mW                      & PC card       \\ \hline
Bluetooth             & 2.4 GHz            & 723.1 kbit/s            & 1 mW                        & PCB module    \\ \hline
Zigbee                & 864 MHz            & 20 kbit/s               & 10 mW                       & PCB module    \\ \hline
Active RFID           & 134 kHz            & 128 bit/s               & \textless 1 mW              & pill          \\ \hline
In vivo communication & \textless 1 MHz    & \textgreater{}64 kbit/s & \textless 1 mW              & band-aid/pill \\ \hline
\end{tabular}}
\end{table}        

In addition, since the \textit{\textit{in vivo }}antennas are radiating into a complex lossy medium, the radiating near fields will strongly couple to the lossy environment. This signifies that the radiated power relies on both the radial and angular positions; hence, the near field effect has to be always taken into account when functioning in an \emph{in vivo} environment \cite{6546986}. The electric and magnetic fields behave differently in the radiating near field compared to the far field. Therefore, the wireless channel inside the body necessitates different link equations \cite{1552452}. It must be noted as well that both the delay spread and  multi-path scattering of a cellular network are not directly applicable to near-field channels inside the body. The reason behind this is the fact that the wavelength of the signal is much longer than the propagation environment in the near field \cite{7032404}.

The authors in \cite{6208476} used an accurate human body to investigate the variation in signal loss at different radio frequencies as a function of position around the body. They noticed significant variations in the Received Signal Strength (RSS) which occur with changing positions of the external receive antenna at a fixed position from the internal antenna \cite{6208476}. Nevertheless, their research did not take into account the basic characterization of the \textit{in vivo} channel. In \cite{455}, the authors used an immersive visualization environment to characterize RF propagation from medical implants. Based on 3-D electromagnetic simulations, an empirical path loss (PL) model is developed in \cite{6025277} to identify losses in homogeneous human tissues.
In \cite{4812182,elayan2016channel,elayan2016vivo,elayan2017bio}, the authors carried out numerical and experimental
investigations of biotelemetry radio channels and wave attenuation
in human subjects with ingested wireless implants{}.

Modeling   the \textit{in vivo} wireless channel including building a phenomenological
path loss model is one of the major research goals in this field. A profound understanding of the channel characteristics is required for defining the channel constraints and the subsequent systems' constraints of a transceiver design \cite{wegmuller2007intra}. 

\newcolumntype{L}{>{\centering\arraybackslash}m{6cm}}

\subsection{Path Loss}

Path loss in \textit{in vivo} channels can be investigated using either a Hertzian dipole antenna or a monopole antenna. The authors in  \cite{7032404}
carried out their study based on Hertzian dipole in which path loss is examined with minimal antenna effects. The length of the Hertzian dipole is so small resulting in   little 
interaction with its surrounding environment. The path loss can be calculated as
\begin{equation}
Path\ Loss(r,\theta,\phi)=20*\log_{10}(\frac{|E|^{}_{r=0}}{|E|^{}_{r,\theta,\phi}})
\end{equation}

where $r$ represents the distance from the origin, i.e. the radius
in spherical coordinates, $\theta$ is the azimuth angle and $\phi$ is the
polar angle. $|E|^{}_{r,\theta,\phi}$ is the the magnitude of the
electric field at the measuring point and $|E|^{}_{r=0}$  is the magnitude of electric field at the origin.

Due to the fact that the \textit{in vivo}
environment is an inhomogeneous medium, it is mandatory to
measure the path loss in the spherical coordinate system \cite{7032404}.
The setup of this approach is depicted in Fig. 5 which includes the truncated human body, the Hertzian dipole, and the spherical coordinate system. \begin{figure}[h]
\centering
\includegraphics[width=0.6\textwidth,clip]{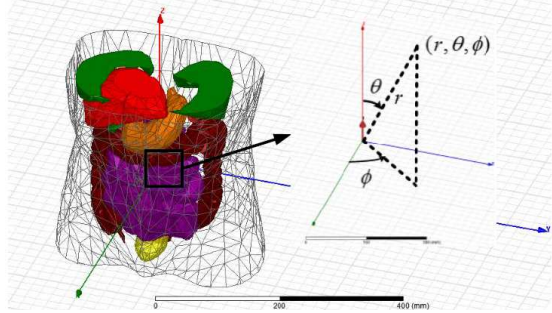}
\caption{\footnotesize Truncated human body with a Hertzian dipole at the origin in spherical coordinate system \cite{7032404}.}
\label{system}
\end{figure}

The authors in  \cite{6208476} carried out their study based on monopole antenna.   Actually, monopoles are good choice of practical antennas since they are small in size, simple and omnidirectional. The path loss can be measured by scattering
parameters (S parameters) that describe the input-output
relationship between ports (or terminals) in an electrical system \cite{6208476}. According to Fig. 6, if we set Port 1 on transmit antenna and Port 2 on receive antenna, then $S_{21}$ represents the power gain of Port 1 to Port 2, that is
\begin{equation}
{|S_{21}|^{2}}=\frac{P_{r}}{P_{t}}
\end{equation}
where $P_{r}$ is the received power and $P_{t}$ is the transmitted power. Therefore, we calculate the path loss by the formula below

\begin{equation}
Path\ Loss(dB)= 20\log_{10}|S_{21}|
\end{equation}
\begin{figure}[h]
\centering
\includegraphics[width=0.6\textwidth,clip]{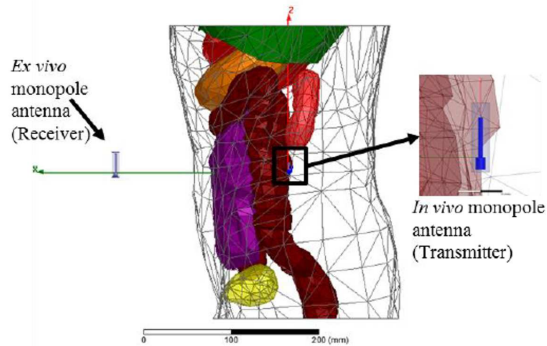}
\caption{\footnotesize Simulation setup by using monopoles to measure the path loss \cite{7032404}.}
\label{system}
\end{figure}

 Based on the simulations presented in \cite{7032404}, it can be observed that there is a substantial difference in the behaviors of the path loss between the \textit{in vivo} and free space environment. In fact, significant attenuation occurs inside the body resulting in an \textit{in vivo} path loss that can be up to 45 dB greater than the free space path loss. Fluctuations in the out-of-body region is experienced by the \textit{in vivo} path loss. On the other hand, free space path loss increases smoothly. The inhomogeneous medium results as well in angular dependent path loss \cite{7032404}.

\subsection{Comparison of \textit{Ex Vivo }and \textit{In Vivo} Channels}

The different characteristics between \textit{ex vivo} and \textit{in vivo} channels are summarized in \cite{7032404} as shown in Table 2.

\begin{table}[h]
\centering
\caption{Comparison of \textit{Ex vivo} and \textit{In vivo} Channel \cite{6857757}}
\label{Simulation}
\scalebox{0.65}{
\begin{tabular}{|L|L|L|}
\hline
Features         & \textit{Ex vivo}                                                                                              & \textit{In vivo}                                                                                                                                                         \\ \hline
Physical Wave Propagtion  & \begin{tabular}[c]{@{}c@{}}Constant speed \\ Multipath - reflection, scattering,\\ and diffraction\end{tabular} & \begin{tabular}[c]{@{}c@{}}Variable speed\\ Multipath and penetration\end{tabular}                                                                                       \\ \hline
Attenuation and Path Loss & \begin{tabular}[c]{@{}c@{}}Lossless medium\\ Decrease inversely with distance\end{tabular}                    & \begin{tabular}[c]{@{}c@{}}Very lossy medium\\ Angular (directional) dependent\end{tabular}                                                                              \\ \hline
Dispersion                & Multipath delays-time dispersion                                                                              & \begin{tabular}[c]{@{}c@{}}Multipath delays of variable speed\\ frequency dependency\\ time dispersion\end{tabular}                                                      \\ \hline
Directionality            & Propagation essentially uniform                                                                               & \begin{tabular}[c]{@{}c@{}}Propagation varies with direction\\ Directionality of antennas \\changes with position\end{tabular}                                             \\ \hline
Near Field Communication  & Deterministic near-field region around the antenna                                                            & \begin{tabular}[c]{@{}c@{}}Inhomogeneous medium - near field\\ region changes with angles \\and position inside the body\end{tabular}                                       \\ \hline
Power Limitations         & Average and Peak                                                                                              & Plus specific absorption rate (SAR)                                                                                                                                      \\ \hline
Shadowing                 & Follows a log normal distribution                                                                             & To be determinted                                                                                                                                                        \\ \hline
Multipath Fading          & Flat fading and frequency selective fading                                                                    & To be determinted                                                                                                                                                        \\ \hline
Antenna Gains             & Constant                                                                                                      & \begin{tabular}[c]{@{}c@{}}Angular and positional dependent\\ Gains highly attenuated\end{tabular}                                                                       \\ \hline
Wavelength                & The speed of light in free space divided by frequency                                                         & \begin{tabular}[c]{@{}c@{}}at 2.4GHz, average dielectric constant \\$\epsilon_r = 35$, which is roughly 6 times \\smaller than the wavelength in \\free space.\end{tabular} \\ \hline
\end{tabular}}
\end{table}

\section{MIMO \textit{In Vivo}}
Due to the lossy nature of the \textit{in vivo} medium, attaining high data rates with reliable performance is considered a challenge \cite{7032380}. The reason behind this is that the \textit{in vivo} antenna performance may  be affected by near-field coupling as mentioned earlier and the signals level will be limited by a specified Specific Absorption Rate (SAR) levels. The SAR is a measurement of how much power is absorbed per unit mass of conductive material, in our case, the human organs \cite{6572751}. This measurement is   limited by the Federal Communications Commission (FCC) which in turns limits the transmission power \cite{6572751}. 

\subsection{Capacity of MIMO \textit{In Vivo} }
The MIMO \textit{in vivo }system capacity is the upper theoretical performance limit that can be achieved in practical systems, and can provide insight into how well the system can perform theoretically and give guidance on how to optimize the MIMO \textit{in vivo }system \cite{6857757}. The achievable transmission rates in the \textit{in vivo} environment have been simulated using a model based on the IEEE 802.11n standard \cite{5307322} because this OFDM-based standard supports up to 4 spatial streams (4x4 MIMO). Owing to the form factor constraint inside the human body,  current studies are restricted to 2x2 MIMO.

The OFDM system can be modeled as:
\begin{equation}
Y_{k}=H_{k}X_{k}+ W_{k},  k=1,2,...N_{data}
\end{equation}

where $Y_{k},X_{k},W_{k} \in C^{2}$ denote the received signal, transmitted signal, and white Gaussian noise with power density of $N_{0}$ respectively at OFDM subcarrier $k$. The symbol $N_{data}$ is the total number of subcarriers configured in the system to carry data. The complex frequency channel response matrix at subcarrier $k$ is denoted by $H_{k}$$\in$ $C^{2*2}$.

The SVD (Singular Value Decomposition) of $H_{k}$ is given as:
\begin{equation}
H_{k}=U_{k}\Sigma_{k}V^{H}_{k}
\end{equation}

where $U_{k}$,$V^{H}_{k}$ $\in$ $C^{2*2}$ are unitary matrices, and $\Sigma_{k}$ is the nonnegative diagonal matrix whose diagonal elements are singular values of $\sqrt{\lambda}_{k1}$, $\sqrt{\lambda}_{k2}$, respectively.

The system capacity for subcarrier $k$ is \cite{tse2005fundamentals}:

\begin{equation}
C_{k}=E[\sum_{i=1}^{2}\log_{2} (1+ \frac{\lambda_{ki}P}{2N_{0}BW})]
\end{equation}

where $P$ is the total transmit signal power of the two transmitter antennas, $BW$ is the configured system bandwidth in $Hz$, and $E$ denotes expectation. In this chapter,  only time-invariant Gaussian channels will be considered. Hence, the expectation in the capacity calculation will be ignored.

The total system capacity is calculated as:
 \begin{equation}
C=\frac{1}{T_{sym}}\sum_{k=1}^{N_{data}} C_{k} = (\frac{BW}{N_{total}}+T_{GI})\sum_{k=1}^{N_{data}} C_{k}
\end{equation}

where $Tsym$ is the duration of each OFDM symbol, $N_{total}$ is the total number of subcarriers available in  bandwidth $BW$, and $T_{GI}$ is the guard interval.

\subsection{Results of MIMO \textit{In Vivo} }

The authors in \cite{6857757} analyzed the Bit Error Rate for a MIMO \textit{in vivo} system. By comparing their results to a $2\times\ 2$  SISO \textit{in vivo}, it was evident that significant performance gains can be achieved when using a $2\times\ 2$  MIMO \textit{in vivo.} This setup allows maximum SAR levels to be met which results in the possibility of achieving target data rates up to 100 Mbps if the distance between the transmit (Tx) and receive (Rx) antennas is within 9.5 cm \cite{6572751}. Fig. 7 below demonstrates the simulation setup that shows the locations of the MIMO antennas. Two Tx antennas are placed inside the abdomen while two Rx antennas are placed at different locations inside the body at the same planar height \cite{6857757}.   
\begin{figure}[h]
\centering
\includegraphics[width=0.6\textwidth,clip]{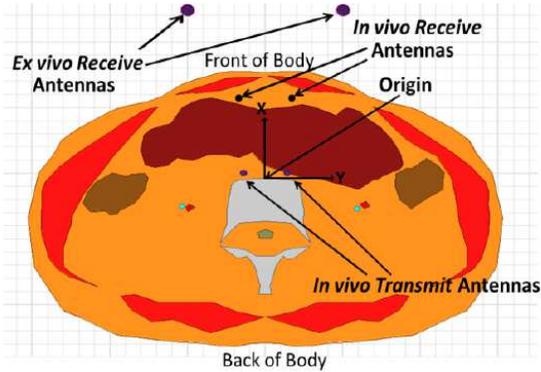}
\caption{\footnotesize Simulation setup showing locations of MIMO antenna \cite{6857757}.}
\label{system}
\end{figure}

The antennas used in Fig. 7 are monopole antennas designed to operate at the 2.4 GHz ISM band in their respective medium which is either  free space for the \textit{ex vivo }antennas or the internal body for the \textit{in vivo} antennas \cite{6857757}. For the \textit{in vivo} case, the monopole's performance and radiation pattern varies with  position and orientation inside the body; therefore, the performance of the \textit{in vivo} antenna is strongly dependent on  the antenna type \cite{6208476}\cite{6546986}.  

Further, in \cite{7032380}, it was proved that not only MIMO \textit{in vivo} can achieve better performance  in comparison to SISO systems but also considerably better system capacity can be observed when Rx antennas are placed at the side of the body. Fig. 8 compares the \textit{in vivo }system capacity for front, right side, left side, and back of the body.  In addition, it was noticed  that in order to meet high data rate requirements of up to 100 Mbps with a distance between the Tx and Rx antennas greater than 12 cm for a 20 MHz channel, relay or other similar cooperative networked communications are necessary to be introduced into the WBAN network \cite{7032380}. 

\begin{figure}[h!]
\centering
\includegraphics[width=0.6\textwidth,clip]{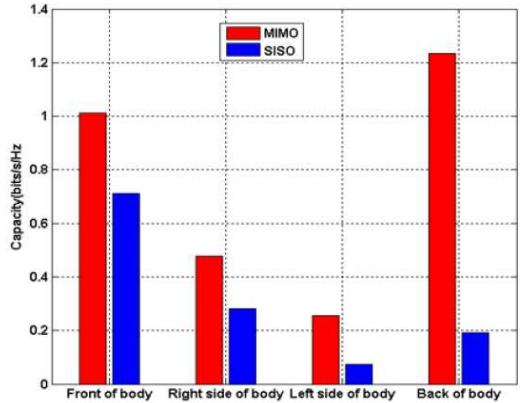}
\caption{ \footnotesize $2\times2$  MIMO and SISO \textit{in vivo} system capacity comparison \cite{7032380}. }
\label{system}
\end{figure}

\subsection{Applications of MIMO \textit{In Vivo}}
One prospective application for MIMO \textit{in vivo }communications is the MARVEL (Miniature Anchored Remote Videoscope for Expedited Laparoscopy) \cite{6225118}. MARVEL is a wireless research platform for advancing MIS (Minimally Invasive Surgery) that necessitates high bit rates ($\sim80$ - $100$ Mbps) for high-definition video transmission with low latency during surgery \cite{6376141} . 

\section{\textit{In Vivo} Nano-communication}

Nanotechnology opens the door towards a new communication paradigm that introduces a variety of novel tools. This technology enables engineers to design and manufacture nanoscale electronic devices and systems with substantially new properties \cite{5450602}. These devices cover radio frequencies in the Terahertz (THz) range and beyond, up to optical frequencies. The interconnections of nanodevices build up into nanonetworks enabling a plethora of potential applications in the biomedical, industrial, environmental and military fields.

\textit{In vivo} nanosensing systems \cite{eckert2013opening}, which can operate inside the human body in real time, have been recently proposed as a way to provide faster and more accurate disease diagnosis
and treatment than traditional technologies based on \textit{in vitro}
medical devices. However, the sensing range of each nanosensor is limited to its close nano-environment; thus, many nanosensors are needed to cover significant regions
or volumes. Moreover, an external device and  user
interaction are necessary to read the actual measurement.
By means of such communication, nanosensors will be able to
overcome their limitations and expand their applications \cite{akyildiz2010electromagnetic}.
Indeed, nanosensors will be able to transmit their information in a multi-hop fashion to a gateway or
sink, react to instructions from a command center, or coordinate
between them in case that a joint response to an event or
remote command is needed.  For instance,  Fig. 9 shows several
\textit{in vivo} nanosensors that communicate as they travel through a
blood vessel. 

A number of challenges exist in the creation of \textit{in vivo }nanosensor networks, which range from the development of nano-antennas for \textit{in vivo} operation to the characterization of the intra-body channel environment from the nanosensor perspective \cite{5381692}.

\begin{figure}[h]
\centering
\includegraphics[width=0.6\textwidth,clip]{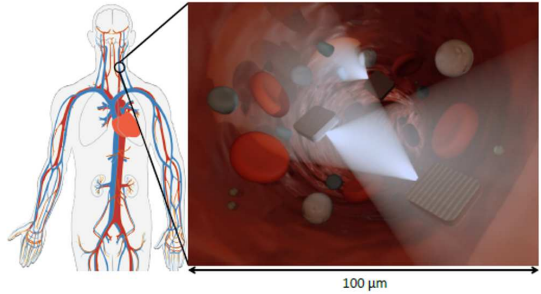}
\caption{\footnotesize \textit{In vivo} nanosensor network inside a blood vessel \cite{5381692}. }
\label{system}
\end{figure}

In order to develop \textit{in vivo }wireless nanosensor networks (iWNSNs), plasmonic
nano-antennas for intra-body communication must be utilized \cite{park2009optical}. In addition, a new view on intra-body channel modeling must
be presented. In traditional channel models, the human body
is modeled as a layered material with different permeabilities
and permittivities. However, from the nanosensor perspective,
and when operating at very high frequencies, the body
is a collection of different elements (cells, organelles and
proteins, among others), with different geometry and arrangement,
as well as different electrical and optical properties.
Further, coupling and interference effects among multiple
nanosensors must  be investigated and utilized at the basis of
novel protocols for iWNSNs. The very high density of
nanosensors in the envisioned applications results in non-negligible
interference effects as well as electromagnetic
coupling among nano-devices \cite{5381692}. 

The future vision of \textit{in vivo} networks entails the distribution of nano-machines that will patrol in the body, take measurements wherever necessary, and send collected data to the outside \cite{akyildiz2014terahertz}. As a result, the development process and the operation of these devices invoke careful measures and high requirements. Moreover, it is important to understand in-body propagation at THz, since it is regarded as the most promising band for electromagnetic paradigm of nano-communication. Actually, the THz Band (0.1-–10 THz) is envisioned as a key technology that satisfies the increasing demand for higher speed wireless communication. THz communication alleviates the spectrum scarcity and capacity limitations of current wireless systems, and hence enables new applications both in classical networking domains as well as in novel nanoscale communication paradigms. Nevertheless, a number communication challenges exist when operating at the THz frequency such as propagation modeling, capacity analysis, modulation schemes, and other physical and link layer design metrics \cite{akyildiz2014terahertz}.

\section{Conclusion}
 This chapter provided an overview of the\textit{ in vivo} communication and networking. The overview focuses on the state of art of the \textit{in vivo }communication, the \textit{in vivo} channel modeling
and characterization, and the concept of MIMO \textit{in vivo}. The
chapter also addresses \textit{in vivo} nano communication which is considered a novel communication paradigm that is going to revolutionize the concept of wireless body area networks. However, several challenges exist which open the door towards further research in this genuine field. 
\footnotesize
\bibliographystyle{IEEEtran}
\bibliography{RefSetPIRMC20132}

\begin{thebibliography}{10}
\providecommand{\url}[1]{#1}
\csname url@samestyle\endcsname
\providecommand{\newblock}{\relax}
\providecommand{\bibinfo}[2]{#2}
\providecommand{\BIBentrySTDinterwordspacing}{\spaceskip=0pt\relax}
\providecommand{\BIBentryALTinterwordstretchfactor}{4}
\providecommand{\BIBentryALTinterwordspacing}{\spaceskip=\fontdimen2\font plus
\BIBentryALTinterwordstretchfactor\fontdimen3\font minus
  \fontdimen4\font\relax}
\providecommand{\BIBforeignlanguage}[2]{{%
\expandafter\ifx\csname l@#1\endcsname\relax
\typeout{** WARNING: IEEEtran.bst: No hyphenation pattern has been}%
\typeout{** loaded for the language `#1'. Using the pattern for}%
\typeout{** the default language instead.}%
\else
\language=\csname l@#1\endcsname
\fi
#2}}
\providecommand{\BIBdecl}{\relax}
\BIBdecl

\bibitem{5931518}
P.~Honeine, F.~Mourad, M.~Kallas, H.~Snoussi, H.~Amoud, and C.~Francis,
  ``Wireless sensor networks in biomedical: Body area networks,'' in
  \emph{Systems, Signal Processing and their Applications (WOSSPA), 2011 7th
  International Workshop on}, May 2011, pp. 388--391.

\bibitem{1632650}
D.~Cypher, N.~Chevrollier, N.~Montavont, and N.~Golmie, ``Prevailing over wires
  in healthcare environments: benefits and challenges,'' \emph{Communications
  Magazine, IEEE}, vol.~44, no.~4, pp. 56--63, April 2006.

\bibitem{shubair2015vivo}
R.~M. Shubair and H.~Elayan, ``In vivo wireless body communications:
  State-of-the-art and future directions,'' in \emph{Antennas \& Propagation
  Conference (LAPC), 2015 Loughborough}.\hskip 1em plus 0.5em minus 0.4em\relax
  IEEE, 2015, pp. 1--5.

\bibitem{shubair_robust_2005}
R.~M. Shubair, ``Robust adaptive beamforming using {LMS} algorithm with {SMI}
  initialization,'' in \emph{2005 {IEEE} {Antennas} and {Propagation} {Society}
  {International} {Symposium}}, vol.~4A, Jul. 2005, pp. 2--5 vol. 4A.

\bibitem{shubair_performance_2005}
R.~M. Shubair and W.~Jessmi, ``Performance analysis of {SMI} adaptive
  beamforming arrays for smart antenna systems,'' in \emph{2005 {IEEE}
  {Antennas} and {Propagation} {Society} {International} {Symposium}}, vol.~1B,
  2005, pp. 311--314 vol. 1B.

\bibitem{khan_compact_2017}
M.~S. Khan, A.~D. Capobianco, S.~M. Asif, D.~E. Anagnostou, R.~M. Shubair, and
  B.~D. Braaten, ``A {Compact} {CSRR}-{Enabled} {UWB} {Diversity} {Antenna},''
  \emph{IEEE Antennas and Wireless Propagation Letters}, vol.~16, pp. 808--812,
  2017.

\bibitem{belhoul_modelling_2003}
F.~A. Belhoul, R.~M. Shubair, and M.~E. Ai-Mualla, ``Modelling and performance
  analysis of {DOA} estimation in adaptive signal processing arrays,'' in
  \emph{10th {IEEE} {International} {Conference} on {Electronics}, {Circuits}
  and {Systems}, 2003. {ICECS} 2003. {Proceedings} of the 2003}, vol.~1, Dec.
  2003, pp. 340--343 Vol.1.

\bibitem{shubair_robust_2004}
R.~M. Shubair and A.~Al-Merri, ``Robust algorithms for direction finding and
  adaptive beamforming: performance and optimization,'' in \emph{The 2004 47th
  {Midwest} {Symposium} on {Circuits} and {Systems}, 2004. {MWSCAS} '04},
  vol.~2, Jul. 2004, pp. II--589--II--592 vol.2.

\bibitem{shubair_closed-form_1993}
R.~M. Shubair and Y.~L. Chow, ``A closed-form solution of vertical dipole
  antennas above a dielectric half-space,'' \emph{IEEE Transactions on Antennas
  and Propagation}, vol.~41, no.~12, pp. 1737--1741, Dec. 1993.

\bibitem{omar_uwb_2016}
A.~Omar and R.~Shubair, ``{UWB} coplanar waveguide-fed-coplanar strips spiral
  antenna,'' in \emph{2016 10th {European} {Conference} on {Antennas} and
  {Propagation} ({EuCAP})}, Apr. 2016, pp. 1--2.

\bibitem{al-ardi_direction_2006}
E.~Al-Ardi, R.~Shubair, and M.~Al-Mualla, ``Direction of arrival estimation in
  a multipath environment: {An} overview and a new contribution,'' in
  \emph{{ACES}}, vol.~21, 2006.

\bibitem{nwalozie_simple_2013}
G.~Nwalozie, V.~Okorogu, S.~Maduadichie, and A.~Adenola, ``A simple comparative
  evaluation of adaptive beam forming algorithms,'' \emph{International Journal
  of Engineering and Innovative Technology (IJEIT)}, vol.~2, no.~7, 2013.

\bibitem{al-nuaimi_direction_2005}
M.~A. Al-Nuaimi, R.~M. Shubair, and K.~O. Al-Midfa, ``Direction of arrival
  estimation in wireless mobile communications using minimum variance
  distortionless response,'' in \emph{Second {International} {Conference} on
  {Innovations} in {Information} {Technology} ({IIT}'05)}, 2005, pp. 1--5.

\bibitem{bakhar_eigen_2009}
M.~Bakhar and D.~P. Hunagund, ``Eigen structure based direction of arrival
  estimation algorithms for smart antenna systems,'' \emph{IJCSNS International
  Journal of Computer Science and Network Security}, vol.~9, no.~11, pp.
  96--100, 2009.

\bibitem{6208476}
T.~Ketterl, G.~Arrobo, A.~Sahin, T.~Tillman, H.~Arslan, and R.~Gitlin, ``In
  vivo wireless communication channels,'' in \emph{Wireless and Microwave
  Technology Conference (WAMICON), 2012 IEEE 13th Annual}, April 2012, pp.
  1--3.

\bibitem{wegmuller2007intra}
M.~S. Wegm{\"u}ller \emph{et~al.}, ``Intra-body communication for biomedical
  sensor networks,'' Ph.D. dissertation, Diss., Eidgen{\"o}ssische Technische
  Hochschule ETH Z{\"u}rich, Nr. 17323, 2007, 2007.

\bibitem{7032380}
C.~He, Y.~Liu, T.~Ketterl, G.~Arrobo, and R.~Gitlin, ``Performance evaluation
  for mimo in vivo wban systems,'' in \emph{RF and Wireless Technologies for
  Biomedical and Healthcare Applications (IMWS-Bio), 2014 IEEE MTT-S
  International Microwave Workshop Series on}, Dec 2014, pp. 1--3.

\bibitem{6063492}
E.~Piel, A.~Gonzalez-Sanchez, H.-G. Gross, and A.~van Gemund, ``Spectrum-based
  health monitoring for self-adaptive systems,'' in \emph{Self-Adaptive and
  Self-Organizing Systems (SASO), 2011 Fifth IEEE International Conference on},
  Oct 2011, pp. 99--108.

\bibitem{5165947}
E.~Chow, B.~Beier, Y.~Ouyang, W.~Chappell, and P.~Irazoqui, ``High frequency
  transcutaneous transmission using stents configured as a dipole radiator for
  cardiovascular implantable devices,'' in \emph{Microwave Symposium Digest,
  2009. MTT '09. IEEE MTT-S International}, June 2009, pp. 1317--1320.

\bibitem{6090392}
Y.~Sun, A.~Anderson, C.~Castro, B.~Lin, R.~Gitlin, S.~Ross, and A.~Rosemurgy,
  ``Virtually transparent epidermal imagery for laparo-endoscopic single-site
  surgery,'' in \emph{Engineering in Medicine and Biology Society, EMBC, 2011
  Annual International Conference of the IEEE}, Aug 2011, pp. 2107--2110.

\bibitem{Zimmerman1995}
T.~G. Zimmerman, ``Personal area networks (pan): near-field intra-body
  communication,'' Master of Science in Media Arts and Sciences, Massachussetts
  Institute of Technology, 1995.

\bibitem{elayan2017terahertz}
H.~Elayan, R.~M. Shubair, J.~M. Jornet, and P.~Johari, ``Terahertz channel
  model and link budget analysis for intrabody nanoscale communication,''
  \emph{IEEE transactions on nanobioscience}, vol.~16, no.~6, pp. 491--503,
  2017.

\bibitem{elayan2017wireless}
H.~Elayan, R.~M. Shubair, and A.~Kiourti, ``Wireless sensors for medical
  applications: Current status and future challenges,'' in \emph{Antennas and
  Propagation (EUCAP), 2017 11th European Conference on}.\hskip 1em plus 0.5em
  minus 0.4em\relax IEEE, 2017, pp. 2478--2482.

\bibitem{668752}
D.~Lindsey, E.~McKee, M.~Hull, and S.~Howell, ``A new technique for
  transmission of signals from implantable transducers,'' \emph{Biomedical
  Engineering, IEEE Transactions on}, vol.~45, no.~5, pp. 614--619, May 1998.

\bibitem{455}
\BIBentryALTinterwordspacing
K.~Sayrafian-Pou, W.-B. Yang, J.~Hagedorn, J.~Terrill, K.~Yekeh~Yazdandoost,
  and K.~Hamaguchi, ``\BIBforeignlanguage{English}{Channel models for medical
  implant communication},'' \emph{\BIBforeignlanguage{English}{International
  Journal of Wireless Information Networks}}, vol.~17, no. 3-4, pp. 105--112,
  2010. [Online]. Available: \url{http://dx.doi.org/10.1007/s10776-010-0124-y}
\BIBentrySTDinterwordspacing

\bibitem{press_ansoft}
\BIBentryALTinterwordspacing
Ansoft, ``{ANSYS HFSS, 3D Full-wave Electromagnetic Field Simulation}.''
  [Online]. Available: \url{http://www.ansoft.com/products/hf/hfss/}
\BIBentrySTDinterwordspacing

\bibitem{gabriel1996compilation}
C.~Gabriel and S.~Gabriel, ``Compilation of the dielectric properties of body
  tissues at rf and microwave frequencies.'' KING'S COLL LONDON (UNITED
  KINGDOM) DEPT OF, Tech. Rep., 1996.

\bibitem{6857757}
C.~He, Y.~Liu, T.~Ketterl, G.~Arrobo, and R.~Gitlin, ``Mimo in vivo,'' in
  \emph{Wireless and Microwave Technology Conference (WAMICON), 2014 IEEE 15th
  Annual}, June 2014, pp. 1--4.

\bibitem{6546986}
A.~Skrivervik, ``Implantable antennas: The challenge of efficiency,'' in
  \emph{Antennas and Propagation (EuCAP), 2013 7th European Conference on},
  April 2013, pp. 3627--3631.

\bibitem{1552452}
H.~Schantz, ``Near field phase behavior,'' in \emph{Antennas and Propagation
  Society International Symposium, 2005 IEEE}, vol.~3B, July 2005, pp. 134--137
  vol. 3B.

\bibitem{7032404}
Y.~Liu, T.~Ketterl, G.~Arrobo, and R.~Gitlin, ``Modeling the wireless in vivo
  path loss,'' in \emph{RF and Wireless Technologies for Biomedical and
  Healthcare Applications (IMWS-Bio), 2014 IEEE MTT-S International Microwave
  Workshop Series on}, Dec 2014, pp. 1--3.

\bibitem{6025277}
D.~Kurup, W.~Joseph, G.~Vermeeren, and L.~Martens, ``In-body path loss model
  for homogeneous human tissues,'' \emph{Electromagnetic Compatibility, IEEE
  Transactions on}, vol.~54, no.~3, pp. 556--564, JUNE 2012.

\bibitem{4812182}
A.~Alomainy and Y.~Hao, ``Modeling and characterization of biotelemetric radio
  channel from ingested implants considering organ contents,'' \emph{Antennas
  and Propagation, IEEE Transactions on}, vol.~57, no.~4, pp. 999--1005, April
  2009.

\bibitem{elayan2016channel}
H.~Elayan and R.~M. Shubair, ``On channel characterization in human body
  communication for medical monitoring systems,'' in \emph{Antenna Technology
  and Applied Electromagnetics (ANTEM), 2016 17th International Symposium
  on}.\hskip 1em plus 0.5em minus 0.4em\relax IEEE, 2016, pp. 1--2.

\bibitem{elayan2016vivo}
H.~Elayan, R.~M. Shubair, A.~Alomainy, and K.~Yang, ``In-vivo terahertz em
  channel characterization for nano-communications in wbans,'' in
  \emph{Antennas and Propagation (APSURSI), 2016 IEEE International Symposium
  on}.\hskip 1em plus 0.5em minus 0.4em\relax IEEE, 2016, pp. 979--980.

\bibitem{elayan2017bio}
H.~Elayan, R.~M. Shubair, and J.~M. Jornet, ``Bio-electromagnetic thz
  propagation modeling for in-vivo wireless nanosensor networks,'' in
  \emph{Antennas and Propagation (EUCAP), 2017 11th European Conference
  on}.\hskip 1em plus 0.5em minus 0.4em\relax IEEE, 2017, pp. 426--430.

\bibitem{6572751}
T.~Ketterl, G.~Arrobo, and R.~Gitlin, ``Sar and ber evaluation using a
  simulation test bench for in vivo communication at 2.4 ghz,'' in
  \emph{Wireless and Microwave Technology Conference (WAMICON), 2013 IEEE 14th
  Annual}, April 2013, pp. 1--4.

\bibitem{5307322}
``Ieee standard for information technology-- local and metropolitan area
  networks-- specific requirements-- part 11: Wireless lan medium access
  control (mac)and physical layer (phy) specifications amendment 5:
  Enhancements for higher throughput,'' \emph{IEEE Std 802.11n-2009 (Amendment
  to IEEE Std 802.11-2007 as amended by IEEE Std 802.11k-2008, IEEE Std
  802.11r-2008, IEEE Std 802.11y-2008, and IEEE Std 802.11w-2009)}, pp. 1--565,
  Oct 2009.

\bibitem{tse2005fundamentals}
D.~Tse and P.~Viswanath, \emph{Fundamentals of wireless communication}.\hskip
  1em plus 0.5em minus 0.4em\relax Cambridge university press, 2005.

\bibitem{6225118}
C.~Castro, S.~Smith, A.~Alqassis, T.~Ketterl, Y.~Sun, S.~Ross, A.~Rosemurgy,
  P.~Savage, and R.~Gitlin, ``Marvel: A wireless miniature anchored robotic
  videoscope for expedited laparoscopy,'' in \emph{Robotics and Automation
  (ICRA), 2012 IEEE International Conference on}, May 2012, pp. 2926--2931.

\bibitem{6376141}
C.~Castro, A.~Alqassis, S.~Smith, T.~Ketterl, Y.~Sun, S.~Ross, A.~Rosemurgy,
  P.~Savage, and R.~Gitlin, ``A wireless robot for networked laparoscopy,''
  \emph{Biomedical Engineering, IEEE Transactions on}, vol.~60, no.~4, pp.
  930--936, April 2013.

\bibitem{5450602}
P.~Russer and N.~Fichtner, ``Nanoelectronics in radio-frequency technology,''
  \emph{Microwave Magazine, IEEE}, vol.~11, no.~3, pp. 119--135, May 2010.

\bibitem{eckert2013opening}
M.~A. Eckert and W.~Zhao, ``Opening windows on new biology and disease
  mechanisms: development of real-time in vivo sensors,'' \emph{Interface
  focus}, vol.~3, no.~3, p. 20130014, 2013.

\bibitem{akyildiz2010electromagnetic}
I.~F. Akyildiz and J.~M. Jornet, ``Electromagnetic wireless nanosensor
  networks,'' \emph{Nano Communication Networks}, vol.~1, no.~1, pp. 3--19,
  2010.

\bibitem{5381692}
J.~M. Jornet, ``Fundamentals of plasmonic communication for in vivo wireless
  nanosensor networks,'' in \emph{36th Annual International Conference on IEEE
  Engineering in Medicine and Biology Society (EMBC)}, Chicago, IL, USA, 2014.

\bibitem{park2009optical}
Q.-H. Park, ``Optical antennas and plasmonics,'' \emph{Contemporary physics},
  vol.~50, no.~2, pp. 407--423, 2009.

\bibitem{akyildiz2014terahertz}
I.~F. Akyildiz, J.~M. Jornet, and C.~Han, ``Terahertz band: Next frontier for
  wireless communications,'' \emph{Physical Communication}, vol.~12, pp.
  16--32, 2014.

\end{thebibliography}


\end{document}